\documentclass{llncs}

\usepackage{amssymb}
\usepackage{mdwlist}
\usepackage{graphicx}
\usepackage{amsmath}
\usepackage{threeparttable}
\usepackage{booktabs}
\usepackage{enumitem}
\usepackage{url}
\newcommand{\keywords}[1]{\par\addvspace\baselineskip
\noindent\keywordname\enspace\ignorespaces#1}

\begin{document}

\mainmatter  


\title{Security of a biometric identity-based encryption scheme}


%
%
\author{Miaomiao Tian%
\thanks{Corresponding author. E-mail: miaotian@mail.ustc.edu.cn (M. Tian).}%
\and Wei Yang \and Liusheng Huang}

%

\institute{School of Computer Science and Technology, University of Science and Technology of China, Hefei, 230026, China\\
Suzhou Institute for Advanced Study, University of Science and Technology of China, Suzhou, 215123, China\\
}

%
%

\maketitle

\begin{abstract}
Biometric identity-based encryption (Bio-IBE) is a kind of fuzzy identity-based encryption (fuzzy IBE) where a ciphertext encrypted under an identity $w'$ can be decrypted using a secret key corresponding to the identity $w$ which is close to $w'$ as measured by some metric. Recently, Yang et al. proposed a constant-size Bio-IBE scheme and proved that it is secure against adaptive chosen-ciphertext attack (CCA2) in the random oracle model. Unfortunately, in this paper, we will show that their Bio-IBE scheme is even not chosen-plaintext secure. Specifically, user $w$ using his secret key is able to decrypt any ciphertext encrypted under an identity $w'$ even though $w$ is not close to $w'$.
\end{abstract}

\section{Introduction}
To simplify the certificate management in traditional public key infrastructure, Shamir \cite{Sha84} first introduced the concept of identity-based cryptography in 1984. In this scenario, a user's public key is derived from his identity, e.g., his e-mail address, and his secret key is generated by a trusted third party called private key generator (PKG) who has knowledge of a master secret key. In 2001, the first two practical identity-based encryption (IBE) schemes were presented in \cite{BF01} and \cite{CC01}, respectively.

The notion of fuzzy identity-based encryption (fuzzy IBE) was introduced by Sahai and Waters \cite{SW05} in 2005, where each identity is viewed as a set of descriptive attributes. A fuzzy IBE scheme is very similar to a standard IBE scheme except that a ciphertext encrypted under an identity $w'$ can be decrypted using the secret key associated with the identity $w$ which is close to $w'$ as judged by some metric. The error-tolerance property of fuzzy IBE enables biometric attributes to be used in a standard IBE scheme. In 2007, Burnett et al. \cite{B07} proposed the first biometric identity-based signature (Bio-IBS) scheme, where they used biometric information to construct the identity of a user. The first biometric identity-based encryption (Bio-IBE) scheme was proposed by Sarier \cite{S08} in 2008. It absorbed the advantage of Burnett et al.'s Bio-IBS scheme. Subsequently, Sarier \cite{S11} presented an improved Bio-IBE scheme which is secure against a new type of denial of service attack. Recently, Yang et al. \cite{Y2011} presented a constant-size Bio-IBE scheme and proved that it is secure against adaptive chosen-ciphertext attack (CCA2) in the random oracle model. Unfortunately, in this paper, we will show that their scheme is even not chosen-plaintext secure.

The rest of this paper is organized as follows. Section 2 introduces some preliminaries required in this paper. In Section 3, we review Yang et al.'s Bio-IBE scheme. In section 4, we present an attack on their Bio-IBE scheme. Finally, we conclude the paper in Section 5.
\section{Preliminaries}
\subsection{Bilinear pairing}
Let $\mathbb{G}$ and $\mathbb{G}_T$ be two groups with the same prime order $p$. A map $e:\mathbb{G}\times\mathbb{G}\rightarrow\mathbb{G}_T$ is called a bilinear map if it satisfies the following three properties.
\begin{enumerate*}
\item Bilinearity: For all $a,b\in \mathbb{Z}_p$ and $u,v\in\mathbb{G}$, we have $e(u^a,v^b)=e(u,v)^{ab}$.
\item Non-degeneracy: There exists $u,v\in\mathbb{G}$ such that $e(u,v)\neq1$.
\item Computability: There is an efficient algorithm to compute $e(u,v)$ for any $u,v\in\mathbb{G}$.
\end{enumerate*}
\subsection{Biometric identity-based encryption}
As mentioned above, a Bio-IBE scheme is essentially a fuzzy IBE scheme, with the only difference that it uses a set of biometric attributes as a user's identity. Therefore, a Bio-IBE scheme also consists of the following four algorithms \cite{SW05}:
\begin{itemize*}
\item \textbf{Setup:} Given a security parameter $k$, the PKG generates a master secret key $MSK$ and the public parameters $PP$ which contains a threshold $d$. The PKG publishes the public parameters $PP$ and keeps the master key $MSK$ secret.
\item \textbf{Extract:} Given the public parameters $PP$, the master secret key $MSK$ and a user's biometric attribute set $w=(\mu_1,\cdots,\mu_n)$, the PKG generates a secret key $sk_w$ for the user.
\item \textbf{Encrypt:} On input the public parameters $PP$, a message $m$ and a user's biometric attribute set $w'=(\mu'_1,\cdots,\mu'_n)$, it returns a ciphertext $C'$.
\item \textbf{Decrypt:} On input the public parameters $PP$, a secret key $sk_w$ corresponding to the user $w$, and a ciphertext $C'$ encrypted under the set of attributes $w'$, it outputs the message if and only if $|w'\bigcap w|\geq d$.
\end{itemize*}

The security notion for Bio-IBE proposed by Yang et al. \cite{Y2011} is indistinguishability of ciphertext under adaptive chosen ciphertext attack (IND-sID-CCA2). A weaker security notion proposed in \cite{SW05} is indistinguishability of ciphertext under chosen plaintext attack (IND-sID-CPA). Its formal definition is based on the following game played between a challenger $\mathcal{C}$ and an adversary $\mathcal{A}$.
\begin{itemize*}
\item \textbf{Init.} The adversary $\mathcal{A}$ outputs a target attribute set $w'=(\mu'_1,\cdots,\mu'_n)$.
\item \textbf{Setup.} The challenger $\mathcal{C}$ runs the \textbf{Setup} algorithm and sends the system parameters $PP$ to the adversary $\mathcal{A}$.
\item \textbf{Phase 1.} The adversary $\mathcal{A}$ adaptively delivers secret key extraction queries on many attribute sets $w_i$, where $|w'\bigcap w_i|< d$ for all $i$. The challenger $\mathcal{C}$ runs the \textbf{Extract} algorithm to obtain a private key $sk_{w_i}$ for each $w_i$ and sends the result to $\mathcal{A}$.
\item \textbf{Challenge.} The adversary $\mathcal{A}$ submits two equal length messages $m_0$ and $m_1$. The challenger $\mathcal{C}$ picks a random bit $b\in\{0,1\}$ and encrypts $m_b$ under $w'$. Then $\mathcal{C}$ sends the ciphertext to $\mathcal{A}$.
\item \textbf{Phase 2.} The adversary $\mathcal{A}$ issues additional secret key extraction queries as in Phase 1.
\item \textbf{Guess.} The adversary $\mathcal{A}$ outputs a guess $b'$ of $b$ and wins if $b'=b$.
\end{itemize*}

The advantage of an adversary $\mathcal{A}$ in this game is defined as $|Pr[b'=b]-1/2|$.\\
\textbf{Definition 1.} A Bio-IBE scheme is IND-sID-CPA secure if there is no polynomial-time adversary that succeeds in the above game with a non-negligible advantage.
\subsection{Fuzzy Extraction}
Fuzzy extraction process is essential for many Bio-IBE schemes such as \cite{S08,S11,Y2011}. Let $\mathcal{M}=\{0,1\}^k$ be a finite dimensional metric space with a distance function $\textsf{dis}:\mathcal{M}\times \mathcal{M}\longrightarrow Z^+$. An $(\mathcal{M},l,t)$ fuzzy extractor consists of the following two functions \textsf{Gen} and \textsf{Rep}:
\begin{itemize*}
\item \textsf{Gen}: This function takes as input a biometric template $b\in\mathcal{M}$. It outputs an identity $ID\in\{0,1\}^l$ and a public parameter $PAR$. The biometric template $b$ is unique for each user since it is a concatenation of user's biometric attributes.
\item \textsf{Rep}: This function takes as input a biometric template $b'\in\mathcal{M}$ and the public parameter $PAR$. It outputs the identity $ID$ if $\textsf{dis}(b,b')\leq t$. In other words, we can obtain the same identity $ID$ as long as $b'$ is ``close" to $b$.
\end{itemize*}
For two biometric attribute sets $w$ and $w'$, we assume that $\textsf{dis}(b,b')\leq t$ if $|w'\bigcap w|\geq d$ and thus we have $ID=ID'$, where $(b,ID)$ and $(b',ID')$ are extracted from $w$ and $w'$, respectively.
\section{Review of Yang et al.'s Bio-IBE scheme}
Let $\Delta_{i,S}(x)=\prod_{j\in S,j\neq i}\frac{x-j}{i-j}$ denote the Lagrange coefficient for $i\in\mathbb{Z}^*_p$ and a set $S$ of elements in $\mathbb{Z}^*_p$. The Yang et al.'s Bio-IBE \cite{Y2011} is specified as follows.

\textbf{Setup:} Given a security parameter $k$, the PKG does:
\begin{enumerate*}
\item Choose two groups $\mathbb{G}$ and $\mathbb{G}_T$ with the same prime order $p$, a bilinear map $e:\mathbb{G}\times\mathbb{G}\rightarrow\mathbb{G}_T$ and a generator $g$ of $\mathbb{G}$.
\item Select two hash functions $H:b\rightarrow\{0,1\}^*$ and $H_1:\mathbb{Z}^*_p\times\{0,1\}^*\rightarrow\mathbb{Z}^*_p$.
\item Pick $s\in \mathbb{Z}^*_p$ and $g_1\in\mathbb{G}$ uniformly at random, and set $g_2=g^s$.
\item The public parameters are $PP=(\mathbb{G},\mathbb{G}_T,e,g,g_1,g_2,d,H,H_1)$ and the master key is $s$.
\end{enumerate*}

\textbf{Extract:} Given a user's biometric attribute set $w=(\mu_1,\cdots,\mu_n)$, the PKG does:
\begin{enumerate*}
\item Compute $ID=H(b)$ and $PAR=\textsf{Gen}(b)$, where $b$ is a concatenation of each $\mu_i$ $(1\leq i\leq n)$.
\item Choose a random $d-1$ degree polynomial $q(x)\in\mathbb{Z}^*_p[x]$ such that $q(0)=s$.
\item For each $i\in[n]$, compute $d_{i,1}=(g_1\cdot g^{H_1(ID)})^{q(\mu_i)}$ and $d_{i,2}=g^{q(\mu_i)}$.
\item Send the private key $sk_w=(d_{i,1},d_{i,2})_{\mu_i\in w}$ to the user and publish $PAR$.
\end{enumerate*}

\textbf{Encrypt:} On input the public parameters $PP$, a message $m\in\mathbb{G}_T$ and an identity $w'=(\mu'_1,\cdots,\mu'_n)$, the sender does:
\begin{enumerate*}
\item Get the public parameter $PAR$ of the receiver and compute $ID'=\textsf{Rep}(b',PAR)$, where $b'$ is a concatenation of each $\mu'_i$ $(1\leq i\leq n)$.
\item Choose $r\in \mathbb{Z}^*_p$ uniformly at random.
\item Compute $C_1=g^r$, $C_2=(g^{H_1(ID')})^r$ and $C_3=m\cdot e(g_1,g_2)^r$.
\item Send $C'=(w',C_1,C_2,C_3)$.
\end{enumerate*}

\textbf{Decrypt:} To decrypt the ciphertext $C'$ encrypted under the attribute set $w'$, a user with attribute set $w$ satisfying $|w'\bigcap w|\geq d$ does:
\begin{enumerate*}
\item Choose an arbitrary set $S\subseteq w'\bigcap w$ such that $|S|=d$.
\item Compute $m=C_3\cdot \frac{e\big(C_2,\prod_{\mu_i\in S}(d_{i,2})^{\Delta_{\mu_i,S}(0)}\big)}{e\big(C_1,\prod_{\mu_i\in S}(d_{i,1})^{\Delta_{\mu_i,S}(0)}\big)}$.
\end{enumerate*}

The \textbf{Decrypt} algorithm works since $ID=ID'$ when $|w'\bigcap w|\geq d$ and
       \begin{eqnarray*}
       &&C_3\cdot\frac{e\big(C_2,\prod_{\mu_i\in S}(d_{i,2})^{\Delta_{\mu_i,S}(0)}\big)}{e\big(C_1,\prod_{\mu_i\in S}(d_{i,1})^{\Delta_{\mu_i,S}(0)}\big)}   \\
       &=&C_3\cdot\frac{e\big((g^{H_1(ID')})^r,\prod_{\mu_i\in S}(g^{q(\mu_i)})^{\Delta_{\mu_i,S}(0)}\big)}{e\big(g^r,\prod_{\mu_i\in S}(g_1\cdot g^{H_1(ID)})^{q(\mu_i)\cdot\Delta_{\mu_i,S}(0)}\big)}  \\
       &=&C_3\cdot\frac{e\big(g^{H_1(ID')\cdot r},g^s\big)}{e\big(g^r,(g_1\cdot g^{H_1(ID)})^s\big)}
       \\
       &=&m\cdot e(g_1,g_2)^r\cdot\frac{e\big(g^{H_1(ID)\cdot r},g^s\big)}{e\big(g^s,(g_1\cdot g^{H_1(ID)})^r\big)}  \\
       &=&m\cdot e(g_1,g^s)^r/e\big(g^s,(g_1)^r\big)  \\
       &=&m
       \end{eqnarray*}
\textbf{Remark.} Compared to the scheme in \cite{Y2011}, there is a small (but important) modification in the above scheme. Namely, we use $H_1(ID)$ (resp. $H_1(ID')$) instead of $H_1(w,ID)$ (resp. $H_1(w',ID')$). We know that, for two random strings $w$ and $w'$, $H_1(w,ID)=H_1(w',ID)$ cannot be true in general. Therefore, the original \textbf{Decrypt} algorithm in \cite{Y2011} may fail. In our modified scheme, the \textbf{Decrypt} algorithm will work since $H_1(ID)=H_1(ID')$ when $|w'\bigcap w|\geq d$. In fact, $H_1(ID)$ plays the same role as $H_1(w,ID)$ in this scheme.
\section{The proposed attack}
Yang et al. \cite{Y2011} proved that their scheme is IND-sID-CCA2 secure in the random oracle model. However, in this section, we show that their scheme is even not IND-sID-CPA secure. Assume that the target attribute set is $w'=(\mu'_1,\cdots,\mu'_n)$. A polynomial time adversary $\mathcal{A}$ attacks Yang at al.'s Bio-IBE scheme as follows:
\begin{enumerate*}
\item In the Setup phase, the adversary $\mathcal{A}$ obtains the system parameters $PP$ from a challenger $\mathcal{C}$.
\item In Phase 1, the adversary $\mathcal{A}$ makes a secret key extraction query on an attribute set $w$, where $|w'\bigcap w|< d$. The challenger $\mathcal{C}$ runs the \textbf{Extract} algorithm to obtain a private key $sk_{w}$ for $w$ and sends the result to $\mathcal{A}$.
\item In Challenge phase, $\mathcal{A}$ submits two equal length messages $m_0$ and $m_1$. The challenger $\mathcal{C}$ picks a random bit $b\in\{0,1\}$ and runs algorithm \textbf{Encrypt$(m_b,w')$} to obtain a ciphertext $C'_b$. Then $\mathcal{C}$ sends $C'_b$ to $\mathcal{A}$.
\item In Phase 2, $\mathcal{A}$ does not issue any query.
\item Let $sk_w=(d_{i,1},d_{i,2})_{\mu_i\in w}=((g_1\cdot g^{H_1(ID)})^{q(\mu_i)},g^{q(\mu_i)})_{\mu_i\in w}$. Upon receiving the ciphertext $C'_b=(w',C_1,C_2,C_3)=(w',g^r,(g^{H_1(ID')})^r,m_b\cdot e(g_1,g_2)^r)$, $\mathcal{A}$ determines the bit $b$ by performing the following steps:
    \begin{enumerate*}
    \item For each $\mu_i\in w$, compute $g_1^{q(\mu_i)}=d_{i,1}/d_{i,2}^{H_1(ID)}$.
    \item Select an arbitrary set $S\subseteq w$ such that $|S|=d$.
    \item Output $m_b=C_3/(\prod_{\mu_i\in S}e(C_1,g_1^{q(\mu_i)})^{\Delta_{\mu_i,S}(0)})$.
    \end{enumerate*}
\end{enumerate*}
We can verify its correctness as follows:
       \begin{eqnarray*}
       &&\frac{C_3}{\prod_{\mu_i\in S}e(C_1,g_1^{q(\mu_i)})^{\Delta_{\mu_i,S}(0)}}  \\
       &=&\frac{m_b\cdot e(g_1,g_2)^r}{\prod_{\mu_i\in S}e(g^r,g_1^{q(\mu_i)})^{\Delta_{\mu_i,S}(0)}}  \\
       &=&\frac{m_b\cdot e(g_1,g_2)^r}{e(g^r,g_1)^s}   \\
       &=&\frac{m_b\cdot e(g_1,g^s)^r}{e(g_1,g^r)^s}      \\
       &=&m_b
       \end{eqnarray*}

It's clear that Yang et al.'s Bio-IBE scheme is broken. That is their scheme is not chosen-plaintext secure. Notice that, in a Bio-IBE scheme, a user with identity $w$ of course can decrypt ciphertexts encrypted under identity $w'$ using his secret key if $|w'\bigcap w|\geq d$. Form the above attack, we know that a user with identity $w$ can also decrypt ciphertexts encrypted under identity $w'$ using his secret key even though $|w'\bigcap w|< d$. Consequently, a valid user can decrypt any ciphertext encrypted under any identity using his secret key in Yang et al. scheme.
\section{Conclusion}
Recently, Yang et al. \cite{Y2011} proposed a constant-size Bio-IBE scheme and proved that it is adaptively chosen-ciphertext secure in the random oracle model. In this paper, however, we have indicated that their scheme is even not chosen-plaintext secure.

\section*{Acknowledgements}
This work was supported by the Major Research Plan of the National Natural Science Foundation of China (No. 90818005), the National Natural Science Foundation of China (No. 60903217), the Fundamental Research Funds for the Central Universities (No. WK0110000027), and the Natural Science Foundation of Jiangsu Province of China (No. BK2011357).







\begin{thebibliography}{00}
\bibitem{Sha84} A. Shamir, ``Identity-based cryptosystems and signature schemes," \emph{Advances in Cryptology (CRYPTO'84)}, LNCS 196, Springer-Verlag, pp.47--53, 1985.
\bibitem{BF01} D. Boneh and M. Franklin, ``Identity-based encryption from the Weil Pairing," \emph{Advances in Cryptology (CRYPTO 2001)}, LNCS 2139, Springer-Verlag, pp.213--229, 2001.
\bibitem{CC01} C. Cocks, ``An identity based encryption scheme based on quadratic residues," \emph{Proceedings of the 8th IMA International Conference on Cryptography and Coding}, LNCS 2260, Springer-Verlag, pp.360--363, 2001.
\bibitem{SW05} A. Sahai and B. Waters, ``Fuzzy Identity-Based Encryption," \emph{Advances in Cryptology (EUROCRYPT 2005)}, LNCS 3494, Springer-Verlag, pp.457--473, 2005.
\bibitem{B07} A. Burnett, F. Byrne, T. Dowling, and A. Duffy, ``A biometric identity based signature scheme," \emph{International Journal of Network Security}, vol. 5, no. 3, pp.317-326, 2007.
\bibitem{S08} N.D. Sarier, ``A New Biometric Identity Based Encryption Scheme," \emph{Proceedings of the 9th International Conference for Young Computer Scientists}, pp.2061-2066, 2008.
\bibitem{S11} N.D. Sarier, ``A new biometric identity based encryption scheme secure against dos attacks," \emph{Security and Communication Networks}, vol. 4, no. 1, pp.23--32, 2011.
\bibitem{Y2011} Y. Yang, Y. Hu, L. Zhang, and C. Sun, ``CCA2 secure biometric identity based encryption with constant-size ciphertext," \emph{Journal of Zhejiang University-SCIENCE C (Computers and Electronics)}, vol. 12, no. 10, pp.819--827, 2011.
\end{thebibliography}
\end{document}